\documentstyle[prl,aps,multicol]{revtex}
\newenvironment{Fig}{
\begin{figure}[h]
\noindent\begin{minipage}[t]{3.4in}
\begin{center}
\leavevmode
\epsfxsize=3 true in
}
{
\end{center}
\end{minipage}
\end{figure}
}
\begin{document}
\bibliographystyle{prsty}
\input epsf
\title{Viscoelasticity and Surface Tension at the Defect-Induced First-Order
Melting Transition of a Vortex Lattice}
\author{Herv\'e M. Carruzzo and Clare C. Yu} 
\address{
Department of Physics and Astronomy, University of California, 
Irvine, Irvine, California 92697}
\date{\today}
\maketitle
\begin{abstract}
We show that thermally activated
interstitial and vacancy defects can lead to first order melting
of a vortex lattice. We obtain good agreement with experimentally
measured melting curve, latent heat, and magnetization jumps for
YBa$_{2}$Cu$_{3}$O$_{7-\delta}$ and Bi$_{2}$Sr$_{2}$CaCu$_{2}$O$_{8}$. 
The shear modulus of the vortex liquid is frequency
dependent and crosses over from zero at low frequencies to a finite
value at high frequencies. We also find a small surface tension
between the vortex line liquid and the vortex lattice.
\end{abstract}


\begin{multicols}{2}
\narrowtext 
\section{\bf Introduction}
It has been experimentally established that below a critical value of the
magnetic field,
vortex lattices undergo a first order transition in clean high temperature
superconductors \cite{Blatter94,Brandt95}. This has been seen in 
YBa$_{2}$Cu$_{3}$O$_{7-\delta}$ (YBCO) 
\cite{Safar93,Liang96,Welp96,Schilling96,Schilling97a,Schilling97b,Roulin98}, 
Bi$_{2}$Sr$_{2}$CaCu$_{2}$O$_{8}$ (BSCCO) 
\cite{Zeldov95,Fuchs96,Keener97} and 
(La$_{1-x}$Sr$_x$)$_2$CuO$_4$ \cite{Sasagawa98}. 
Evidence for first order phase transitions 
comes from latent heat measurements
\cite{Schilling96} and peaks in the specific heat
\cite{Schilling97a,Schilling97b,Roulin98}as well as jumps in the resistivity 
\cite{Safar93,Fuchs96,Keener97} 
and in the magnetization \cite{Liang96,Welp96,Zeldov95,Sasagawa98}. 
This transition is generally accepted as a melting transition
from a vortex solid to a vortex liquid. 

A great deal of theoretical work
\cite{Brezin85,Houghton89,Hetzel92,Sasik95,Ryu96,Ryu97,Hu97,Hu98,Nguyen96,Nguyen98,Nordborg97}
has helped to establish that there is a first order melting
transition. Br\'ezin, Nelson and
Thiaville \cite{Brezin85}
showed that including fluctuation effects in Abrikosov's
mean field theory of the flux lattice transition
would drive the transition first order. The advent of
the high temperature superconductors and the subsequent
experimental indications of vortex lattice melting sparked intense
theoretical activity. 
Early analytic efforts used the Lindemann criterion \cite{Houghton89}, though
such an approach could not show that the transition was first order.
Studies of the mapping between vortex lines and the world lines of
a 2D system of bosons have suggested a first order transition
from an Abrikosov lattice to an entangled vortex liquid 
\cite{Nelson88,Nelson89,Nordborg97}. Numerical simulations 
\cite{Hetzel92,Sasik95,Ryu96,Ryu97,Hu97,Hu98,Nguyen96,Nguyen98,Li94}
have been able to show that the melting is first
order by calculating quantities such as the magnetization jump
\cite{Sasik95,Ryu97} and the delta function in the specific heat
\cite{Hu97,Hu98,Nguyen98}. However these simulations
were done in the limit of high magnetic fields 
$\xi_{ab}\ll a_o \ll \lambda_{ab}$ where $\xi_{ab}$ is the coherence length,
$a_o$ is the spacing between vortices, and $\lambda_{ab}$ is 
the penetration depth. Most 
\cite{Hetzel92,Sasik95,Ryu97,Hu97,Hu98,Nguyen96,Nguyen98}
assumed that the magnetic induction ${\bf B}$
was spatially uniform and thus neglected the wavevector dependence
of the elastic moduli. This has made quantitative comparison with
experimental data difficult, especially in the case of BSCCO 
whose vortex lattice melts at low fields.
In addition the mechanism for vortex lattice melting is 
still not well established. 
There have been suggestions that topological defects 
\cite{Tesanovic95} such as
vortex loops \cite{Nguyen96,Nguyen98}, vortex--antivortex
pairs \cite{Hu97,Hu98}, free disclinations \cite{Ryu96}, and 
dislocations \cite{Kierfeld99} may 
play a key role in triggering melting.

\subsection{Melting Scenario}
In this paper we show that melting can be induced by
interstitial and vacancy line defects in the
vortex lattice which soften the shear modulus $c_{66}$.
This softening makes it easier to introduce more defects
and increases the vibrational free energy. The 
increased vibrations ultimately lead to melting. 
There is good agreement with the experimental curve of transition
temperature versus field, latent heat and magnetization
jumps for YBCO and BSCCO. Using a viscoelastic
approach, we show that the shear modulus is frequency dependent.
At zero frequency the vortex liquid cannot sustain a shear while
at high frequency the liquid has a finite shear modulus. Since we
can calculate the free energy for both the lattice and the liquid
at melting, we have estimated the surface tension between a vortex
line liquid and a vortex solid.

Let us describe our scenario for melting. 
Our approach follows that of Granato \cite{Granato92}
as well as previous work which showed that defects can
lead to a first order phase transition \cite{Carruzzo98}.
We start with a 
vortex lattice in a clean layered superconductor with a magnetic
field $H$ applied perpendicular to the layers along the
c-axis. We consider the vortices
to be correlated stacks of pancake vortices. 
We will assume that the transition is induced
by topological defect lines, i.e., vacancies and interstitials.
In a Delaunay triangulation \cite{Preparata85}
a vacancy or an interstitial in a triangular lattice
is topologically equivalent to a pair of bound dislocations \cite{Ryu96}
as well as to a twisted bond defect \cite{Kim96}.
High temperature decoration experiments \cite{Kim96}
and Monte Carlo simulations \cite{Ryu96} have found such
defects to be thermally excited. The introduction
of these defects softens the elastic moduli. Since the energy to introduce
interstitials and vacancies is proportional to the elastic moduli, softening
makes it easier to introduce more defects. The softening also increases
the vibrational entropy of the vortex lattice which leads to a
melting transition. The transition is driven
by the increased vibrational entropy of the vortex lines
of the lattice, and not by
the entropy of the wandering of the defect lines. In fact
Frey, Nelson and Fisher \cite{Frey94} showed that a phase transition
driven by the entropy of wandering flux lines occurs at a much
higher magnetic field than what is observed experimentally. In the vicinity
of the experimentally observed first order phase transition,
wandering in the transverse direction by more than a lattice spacing
is energetically quite costly and therefore rare. Such flux line bending
also makes dislocations \cite{Marchetti90,Marchetti99}
energetically costly at low dislocation densities. (The energy scale 
is set by $\epsilon_{o}s$ \cite{Blatter94,Brandt95}.
Here $s$ is the interplane spacing and $\epsilon_{o}$, the energy
per unit length of a vortex, is given by 
$\epsilon_{o}=(\phi_{o}/4\pi\lambda_{ab})^2$
where $\phi_{o}$ is the flux quantum and $\lambda_{ab}$ is the penetration
depth for currents in the {\it ab} plane. For example, for YBCO 
$\epsilon_{o}s\sim 650$ K at $T=70$ K 
and for BSCCO $\epsilon_{o}s\sim 550$ K at $T=60$ K \cite{comment:wander}.
Note that $\epsilon_{o}s\gg T$.)

The first order transition is nucleated in a small
region by a local rearrangement of existing line segments. Slightly above the
melting temperature $T_{m}$ a vortex line can distort and make an
interstitial and a vacancy line segment that locally melt the solid. 
This is the
analog of a liquid droplet which nucleates melting of a crystal.
The role of the surface tension is played by the energy to connect
the interstitial segment to the rest of the vortex line.
This connection can be a Josephson vortex lying between planes or a
series of small pancake vortex displacements spread over several layers.
When the length $\ell$ of the interstitial and vacancy segments equals
the critical length $\ell_{c}$, the energy gained
by melting equals the energy cost of the connections. 
When $\ell>\ell_{c}$, it is energetically favorable for the defect
segments grow to the length of the system. We are ignoring the
surface tension associated with the surface parallel to the c--axis.
We shall show later that this is quite small.

\section{\bf Free Energy}
To study melting we assume that we have a vortex lattice with
interstitial and vacancy lines extending the length of the lattice.
Our goal is to find the free energy density as a function of the
concentration $n$ of defect lines.
The free energy density is 
\begin{equation}
f=f_{o}+f_{w}+f_{vib}+f_{wan}
\label{eq:freeng}
\end{equation}
where $f_{o}$ is the free energy density of a perfect lattice,
$f_{w}$ is the work needed to introduce a straight interstitial 
or vacancy line into the lattice,
$f_{vib}$ is the vibrational free energy density of the system, and
$f_{wan}$ is the free energy due to the wandering
of the defect lines over distances large compared to the lattice spacing.
We now examine these terms in detail.

$f_{o}$, the free energy density of a perfect rigid 
flux lattice, is given by the London term \cite{Frey94,Tinkham96}:
\begin{equation}
f_{o}=\frac{B^{2}}{8\pi}+\frac{B\phi_{o}}{32\pi^{2}\lambda_{ab}^{2}}
\ln\left(\frac{\eta\phi_{o}}{2\pi\xi_{ab}^{2}B}\right),\;\;
\frac{\phi_{o}}{4\pi\lambda_{ab}^{2}}\ll B\ll H_{c2}
\label{eq:london}
\end{equation}
where 
$B$ is the spatially averaged magnetic induction, 
$\xi_{ab}$ is the
coherence length in the $ab$ plane, and $\eta$ is 0.130519 for
a hexagonal lattice and 0.133311 for a square lattice \cite{Frey94}.
For $B$ near $H_{c2}$, $f_{o}$ is given by
the Abrikosov free energy \cite{Degennes89}
\begin{equation}
f_{o}=\frac{B^{2}}{8\pi}-\frac{(H_{c2}-B)^{2}}
{8\pi[1+(2\kappa^{2}-1)\beta_{A}]}
\label{eq:abrikosov} 
\end{equation}
where the Ginzburg-Landau parameter $\kappa=\lambda_{ab}/\xi_{ab}$,
and the Abrikosov parameter $\beta_{A}$ is 1.16 for a triangular lattice
and 1.18 for a square lattice.

To calculate $f_{vib}$, we follow ref. \cite{Bulaevskii92}. 
We denote the displacement
of the $\nu$th vortex pancake in the $n$th plane from its
equilibrium position by ${\bf u}(n,\bf{r}_{\nu})$ where
${\bf u}=(u_{x},u_{y})$ and the pancake position 
${\bf r}=(r_{x},r_{y})$. The Fourier transform
${\bf u}({\bf k},q)=\sum_{n \nu}\bf{u}(n,\bf{r}_{\nu})\exp[i({\bf k}\cdot
{\bf r}_{\nu} + qn)]$. $\bf{k}=(k_{x},k_{y})$ and $q$ is the wavevector
along the c-axis. $f_{vib}=-(k_{B}T/V)\ln Z_{vib}$ where $V$ is the
volume and the vibrational partition function $Z_{vib}$ is given by
\begin{equation}
Z_{vib} = \int e^{-{\cal F}_{el}/k_{B}T}
\prod_{\bf{k},q>0,i}
\frac {du_{R}(i {\bf k},q)du_{I}(i {\bf k},q)}
{\pi\xi_{ab}^2} 
\label{eq:Z}
\end{equation}
where we have divided by the area $\pi\xi_{ab}^2$ of the normal
core of a pancake \cite{Bulaevskii92}. $u_{R}$ and $u_{I}$ are the real
and imaginary parts of ${\bf u}(\bf{k},q)$ and $i\epsilon \{x,y\}$.
The elastic free energy functional associated with these distortions is 
\begin{equation}
{\cal F}_{el} =\frac{1}{2}\upsilon_{o}\sum_{{\bf k}q}
\sum_{ij}u_{i}({\bf k},q)a_{ij}u_{j}^{*}({\bf k},q)
\label{eq:elastic}
\end{equation}
where $i$ and $j\epsilon \{x,y\}$, the volume per
pancake vortex is $\upsilon_{o}=s\phi_{o}/B$, and
$s$ is the interplane spacing. 
The $\bf{k}$ sum is over a circular Brillouin
zone $K_{o}^{2}=4\pi B/\phi_{o}$. The matrix $a_{ij}$ is given by
\begin{equation}
a_{ij}=c_{B}k_{i}k_{j}+(c_{66}k^{2}+c_{44}Q^{2})\delta_{ij}
\label{eq:aij}
\end{equation}
where $c_{B}$, $c_{66}$, and $c_{44}$ are the bulk, shear, and tilt
moduli, respectively. $c_{B}=c_{11}-c_{66}$ for a hexagonal lattice.
$Q^{2}=2(1-\cos qs)/s^{2}$. 
Diagonalizing $a_{ij}$ leads to 2 eigenvalues:
\begin{eqnarray}
A_{\ell}(kq)&=&c_{11}k^{2}+c_{44}Q^{2} \nonumber\\ 
A_{t}(kq)&=&c_{66}k^{2}+c_{44}Q^{2}\\
\label{eq:A}
\end{eqnarray}
where $A$ is the diagonal matrix, the subscript $\ell$ 
denotes longitudinal and $t$ denotes transverse. Using this leads to
\begin{equation}
{\cal F}_{el} =\frac{1}{2}\upsilon_{o}\sum_{{\bf k}q}
\sum_{i=\ell,t}A_{i}|u_{i}({\bf k},q)|^2
\end{equation}
where $i\epsilon \{\ell,t\}$.
After integrating over $u$ in (\ref{eq:Z}),
the remaining sums over $\bf{k}$ and $q$ are converted to 
integrals:
\begin{equation}
\ln Z_{vib} = \sum_{i=\ell,t}\frac{1}{2}\int_{0}^{K^2_o}
\frac{d\left(k^2\right)}{4\pi}\int_{-\pi/s}^{\pi/s}\frac{dq}{2\pi}
\ln\left(\frac{2k_BT}{\upsilon_{o}\xi_{ab}^2 A_i}\right)
\label{eq:Zintegrals}
\end{equation}
where the volume of the sample is set to unity. 
The integrals in (\ref{eq:Zintegrals}) are done numerically.
At low fields ($b=B/H_{c2}<0.25$), the
elastic moduli are given by \cite{Blatter94,Brandt95,Brandt77b}
\begin{eqnarray}
c_{66} & = & \frac{B\phi_{o}\zeta}{(8\pi\lambda_{ab})^{2}} \nonumber \\
c_{11} & = & \frac{B^{2}[1+\lambda_{c}^{2}(k^{2}+Q^{2})]}
{4\pi[1+\lambda_{ab}^{2}(k^{2}+Q^{2})](1+\lambda_{c}^{2}k^{2}
+\lambda_{ab}^{2}Q^{2})} \nonumber \\
c_{44} & = & \frac{B^{2}}{4\pi(1+\lambda_{c}^{2}k^{2}+\lambda_{ab}^{2}Q^{2})}
+ \frac{B\phi_{o}}{32\pi^{2}\lambda_{c}^{2}} \label{eq:moduli} \\
       &   & \mbox{} \times \ln\frac{\xi_{ab}^{-2}}{K_{o}^{2}+(Q/\gamma)^{2}
+\lambda_{c}^{-2}} + \frac{B\phi_{o}}{32\pi^{2}\lambda_{ab}^{4}Q^{2}}
\ln(1+\frac{Q^{2}}{K_{o}^{2}}) \nonumber \\
\nonumber
\end{eqnarray}
where $\lambda_{c}$ is the penetration depth for currents along the
c-axis, $\gamma=\lambda_{c}/\lambda_{ab}$ is the anisotropy, and $\zeta=1$. 
At high fields ($b>0.5$) \cite{Blatter94,Brandt95,Brandt77a}, $c_{66}$ 
is altered by the factor
$\zeta\approx(1-0.5\kappa^{-2})(1-b)^{2}(1-0.58b+0.29b^{2})$
and the penetration depths in $c_{11}$ and $c_{44}$ are replaced by
$\tilde{\lambda}^{2}=\lambda^{2}/(1-b)$ 
where $\lambda$ denotes either $\lambda_{ab}$ or $\lambda_{c}$.
In addition the last two terms of $c_{44}$ are replaced by
$B\phi_{o}/(16\pi^{2}\tilde{\lambda}_{c}^{2})$. These
replacements guarantee that the elastic moduli vanish at $H_{c2}$.
For YBCO
the temperature dependence of the penetration depths and coherence
lengths are given by $\lambda(T)=\lambda(0)(1-(T/T_{c}))^{-1/3}$
\cite{Kamal94} and $\xi_{ab}(T)=\xi_{ab}(0)(1-(T/T_{c}))^{-1/2}$, 
respectively.
For BSCCO whose melting field is two orders 
of magnitude below $H_{c2}$, 
$\lambda^{2}(T)=\lambda^{2}(0)/(1-(T/T_{c})^{4})$ and
$\xi_{ab}^{2}(T)=\xi_{ab}^{2}(0)/(1-(T/T_{c})^{4})$ \cite{Tinkham96}.

The free energy density $f_{w}$ due to the energy cost
of adding a vacancy or interstitial vortex line is difficult
to calculate accurately \cite{Frey94,Olive98}. However, we can
write down a plausible form for $f_{w}$ by noting that a straight
line defect parallel to the c-axis
produces both shear and bulk (but not tilt) distortions
of the vortex lattice. For example, if a defect at the origin produces
a displacement $\bf{u}$ that satisfies 
$\nabla\cdot{\bf u}=\upsilon_{o}\delta({\bf r})/s$ where
$\delta({\bf r})$ is a two dimensional delta function, then
$u_{\alpha}({\bf k})=ik_{\alpha}/k^{2}$ \cite{Frey94,Brandt97}. Inserting
this in (\ref{eq:elastic}), we find that 
$f_{w}=(c_{66}+\overline{c}_{B})/2$ where 
$\overline{c}_{B}=\sum_{\bf{k}}c_{B}(q=0,\bf{k})$. Generalizing
this to allow for a more complicated distortion
and for a concentration $n$ of line defects, we write \cite{Granato92}
\begin{equation}
f_{w}=\int_{0}^{n} dn(\alpha_{1}c_{66}+\alpha_{2}\overline{c}_{B})
\label{eq:fw}
\end{equation}
where $\alpha_{1}$ and $\alpha_{2}$ are dimensionless constants.
We expect the isotropic distortion to be small, i.e., $\alpha_{2}\ll 1$,
and the shear deformation to dominate, i.e., $\alpha_{1}\gg\alpha_{2}$.
Integrating over $n$ allows the elastic moduli to depend
on defect concentration. We will assume that $c_{B}$ is independent of
$n$ since we believe that the bulk modulus of the vortex solid is
roughly the same as that of the liquid phase. To find $c_{66}(n)$
\cite{Granato92}, we use its definition 
\begin{equation}
c_{66}=\partial^{2}f/\partial\varepsilon^{2}
\end{equation}
where $\varepsilon$ is the shear strain. Assuming
that $c_{B}$ has negligible shear strain dependence, we find
\begin{equation}
c_{66}(n)=c_{66}(0)+\alpha_{1}\int_{0}^{n}(\partial^{2}c_{66}(n)/
\partial\varepsilon^{2})dn
\end{equation}
or
\begin{equation}
\frac{\partial c_{66}(n)}{\partial n} = \alpha_{1}\frac{\partial^{2}c_{66}(n)}
{\partial\varepsilon^{2}}
\label{eq:c66}
\end{equation}
If we shear the lattice in the $ab$ plane along rows separated by a distance
$d$, the system must be unchanged if the displacement is equal
to a lattice spacing. This is a result of the discrete translational
symmetry of the lattice. The shear modulus should reflect this discrete
translational symmetry and therefore must be periodic in displacements equal to
the lattice constant $a_{o}=\sqrt{\phi_o/B}$. 
We describe this with the simplest even periodic function: 
\begin{eqnarray}
c_{66}(u)&=&c_{66}(u=0)\cos(2\pi u/a_{o}) \nonumber\\
&=& c_{66}(\varepsilon=0)\cos(2\pi d\varepsilon/a_{o})
\label{eq:cosinec66}
\end{eqnarray}
where the shear strain $\varepsilon=u/d$.
Notice that this expression goes beyond the usual harmonic approximation.
Then taking the second derivative of eq. (\ref{eq:cosinec66}),
we obtain
\begin{equation}
\partial^{2}c_{66}(n)/\partial\varepsilon^{2}=-\beta c_{66}(n)
\end{equation}
where $\beta=4\pi^{2}d^{2}/a_{o}^{2}$. Combining this with
(\ref{eq:c66}), we obtain 
\begin{equation}
c_{66}(n)=c_{66}(0)\exp(-\alpha_{1}\beta n)
\label{eq:c66exp}
\end{equation}
where $c_{66}(0)$ is given in eq. (\ref{eq:moduli}).
Thus the shear modulus softens exponentially with the defect
concentration $n$. This softening lowers the energy cost to introduce
further defects, and increases the vibrational free energy $f_{vib}$
when $c_{66}(n)$ is used in $a_{ij}$. 
Substituting $c_{66}(n)$ in eq. (\ref{eq:c66exp})
into our expression (\ref{eq:fw}) for
$f_{w}$ yields
\begin{equation}
f_{w}=\frac{c_{66}(n=0)}{\beta}(1-e^{-\alpha_{1}\beta n}) +
\alpha_{2}\overline{c}_{B}n
\label{eq:finalfw}
\end{equation}

The softening of the shear modulus with increasing defect concentration
is well known in the case of atomic lattices \cite{Granato94}. 
There it has been shown both experimentally \cite{Holder74a,Holder74b,Rehn74}
and theoretically \cite{Dederichs78} that interstitials can substantially
soften the elastic constants with the largest 
change being in the shear modulus. Linear extrapolation of the
experimentally measured change of the shear modulus of copper would
imply that the lattice becomes unstable for a concentration of about 3\% 
interstitials \cite{Granato94}. 
An example of how interstitials can soften the shear modulus
is illustrated in Figure \ref{fig:dumbbell}.
Here we show a triangular lattice where an interstitial forms a dumbell 
aligned in the $<010>$ direction
by sharing a site with another atom or flux line. Dumbbell displacements
along the $<100>$ direction introduce 
a string--like librational resonance mode consisting of displacements
along the $<110>$ directions. This mode couples strongly to an external
shear stress and results in softening of the shear modulus 
\cite{comment:marchetti}.

The last term we need to consider is $f_{wan}$, the free energy 
due to the wandering
of the defect lines over distances large compared to the lattice spacing.
We can estimate $f_{wan}$ with the following expression \cite{Frey94}
\begin{equation}
f_{wan}\approx -\frac{k_{B}T}{\ell_{z}a_{o}^{2}}\ln(m_{\ell})
\label{eq:wandering}
\end{equation}
where $m_{\ell}=3$ for a triangular lattice (BSCCO) and $m_{\ell}=4$
for a square lattice (YBCO). $\ell_{z}$ can be thought of as the
distance along the z--axis that it takes \cite{Pastoriza95}
for the defect line to wander
a transverse distance of one lattice spacing $a_{o}$. To go from
one vacancy or interstitial site to the next, the defect line segment
must jump over the barrier between the two positions. This gives
$\ell_{z}$ a thermally activated form: 
$\ell_{z}\sim \ell_{o}\exp(-E/k_{B}T)$, where 
$\ell_{o}\approx a_{o}(\epsilon_{1}/\epsilon_{B})^{1/2}$ and
$E\approx a_{o}(\epsilon_{1}\epsilon_{B})^{1/2}$.
$\epsilon_{1}$ is the line tension and is given by
$\epsilon_{1}\sim (\epsilon_{o}/\gamma^{2})\ln(a_{o}/\xi_{ab})$.
Numerical simulations \cite{Frey94,Olive98} indicate that
the barrier height $\epsilon_{B}$ is small and we use
$\epsilon_{B}=2.5\cdot 10^{-3}\epsilon_{o}$.
$f_{wan}$ itself is quite small compared to the other terms because of
the high energy cost of vortex displacements. For example,
at the transition $f_{wan}$ is about
two orders of magnitude smaller than $f_{w}$ or $f_{vib}$.
Thus the transition is not
driven by a proliferation of wandering defect lines because
near the transition the high energy cost of vortex displacements is not
sufficiently offset by the entropy of the meandering line \cite{Frey94}.

Before we plot $f$ versus $n$, we note that the difference
between $B$ and $H$ is negligible for YBCO but can be a significant
fraction of the melting field $H_{m}$ for BSCCO. To find the
value of $B$ to use in the Helmholtz free energy density $f$,
we minimize the Gibbs free energy density $G$, i.e.,
$\partial G/\partial B=0$ where $G=f-{\bf B}\cdot{\bf H}/4\pi$.
Since the concentration dependence of $B$ is negligible, we find
$B$ for $n=0$ for each value of $H$ and $T$. Typical plots of
$\Delta f=f(n)-f(0)=f_{w}+\Delta f_{vib}$ versus $n$ 
are shown in the inset of figure 1. The
double well structure of $\Delta f$ is characteristic of a first
order phase transition. The equilibrium transition occurs 
when both minima have the same
value of $\Delta f$. We associate the minimum at $n=0$
with the vortex solid and the minimum at finite $n$ with the vortex
liquid. The defect concentration at the transition
is only a few percent. Eq. (\ref{eq:c66exp}) implies that a finite
value of $n$ yields a finite value for the shear modulus, e.g, for
$n=5$\%, $c_{66}(n)\sim 0.2c_{66}(0)$ for BSCCO.
Previous work interpreted this to mean that the lattice did not melt.
However, they did not appreciate the fact that the shear modulus is frequency
dependent and the $c_{66}$ used here is the high frequency
response. At high frequencies it is the elastic response which dominates
and this is what enters into the expression for the free energy. For
a liquid the low frequency response is dominated by viscosity so that
the zero frequency shear modulus is zero. 
We will elaborate more on this later.

\section{\bf Fits to Experimental Data}
\subsection{\bf Melting Curve}
In Figure \ref{fig:melting} we fit
the experimental first order transition curves in the $H-T$ plane
using 2 adustable parameters: $\alpha_{1}$ and $\alpha_{2}$.
As expected, $\alpha_{1}\gg\alpha_{2}$ and $\alpha_{2}\ll 1$
(see Figure \ref{fig:melting}). The geometrical 
quantity $\beta$ can have several
values for a given lattice structure, depending on which planes
are sheared. We choose $\beta=\pi^{2}\tan^{2}\phi$ where
$\phi$ is the angle between primitive vectors.
Decoration experiments on BSCCO find a triangular
lattice \cite{Kim96}, so we use $\phi=60^{o}$. 
For YBCO we choose $\phi=44.1^{o}$ which is very close
to a square lattice which has $\phi=45^{o}$. Maki \cite{Won96}
has argued that the $d$-wave symmetry of the order parameter
yields a square vortex lattice tilted by 45$^{o}$ from the
$a-$axis. Experiments \cite{Yethiraj93,Keimer94,Maggio-Aprile95,comment:phi} 
on YBCO find $\phi$ ranging from $36^{o}$ to $45^{o}$. 

\subsection{\bf Magnetization and Entropy Jumps} 
We can calculate the jump in magnetization 
$\Delta M$ at the transition using
$\Delta M = -\partial\Delta G/\partial H|_{T=T_{m}}$ where
$\Delta G=G(n_{\ell})-G(n=0)$. Here $n_{\ell}$ is the defect
concentration in the liquid at the melting transition. The jump
in entropy $\Delta s$ is given by 
$\Delta s=-\upsilon_{o}\partial\Delta G/\partial T|_{H=H_{m}}$ where 
$\Delta s$ is the entropy change per vortex per layer. The results
are shown in Figure \ref{fig:mag-S}. 
We have checked that our results
satisfy the Clausius-Clapeyron equation 
$\Delta s=-(\upsilon_{o}\Delta B/4\pi)dH_{m}/dT$. We obtain
good agreement with experiment for YBCO and the right
order of magnitude for BSCCO. The difference 
between theory and experiment in the temperature
dependence of the entropy and magnetization jumps for BSCCO may be due
to the decoupling of the planes 
\cite{Glazman91,Daemen93,Blatter96,Horovitz98,Dodgson99}. 
This enhances the thermal excursions and hence
the entropy of the pancake vortices \cite{Dodgson98a,Dodgson98b,Rae98}.
Decoupling may be brought about by other
types of defects such as dislocations which we will discuss in section 5B. 

\subsection{\bf Lindemann Criterion}
We can compare our results with the Lindemann criterion by
calculating the mean square displacement $<|u|^{2}>$ at the transition
using eq. (\ref{eq:Z}):
$<|u|^{2}>=-(2k_{B}T/\upsilon_{o})\sum_{\alpha{\bf k}q}
\partial\ln Z_{vib}/\partial A(\alpha{\bf k}q)$ where
$A$ is given by (\ref{eq:A}) and $\alpha$
labels the 2 eigenvalues. Defining the Lindemann ratio $c_{L}$
by $c_{L}^{2}=<|u|^{2}>/a_{o}^{2}$, we find that $c_{L}\approx 0.25$
for YBCO at $H_{m}=5$ T and that $c_{L}\approx 0.11$ for BSCCO at
$H_{m}=200$ G. 
Here we have used the same values of the parameters
that were used to fit the phase transition
curves in Figure \ref{fig:melting}. These
values of $c_{L}$ are consistent with previous values 
\cite{Blatter94,Brandt95,Houghton89}.

\subsection{\bf Hysteresis}
Experiments have found little, if any, hysteresis 
\cite{Safar93,Zeldov95,Keener97}. This is consistent with
our calculations. We can bound 
the hysteresis by noting the range of temperatures between
which the liquid minimum appears and the solid minimum disappears.
Typical values for the width of this temperature 
range are 300 mK for YBCO at $H=5T$ and 1.3 K for
BSCCO at $H=200$ G. Another measure of the hysteresis can be found
in the plots of $\Delta f$ versus $n$ (see inset of Fig. \ref{fig:melting}).
The barrier height $V_{B}$ between the minima
is low ($V_{B}\upsilon_{o}\sim 30$ mK) which is consistent with minimal
hysteresis. 

\subsection{\bf Loss of Superconducting Phase Coherence}
In going from the normal metallic phase to the vortex solid, two
symmetries are broken: translational invariance and gauge
symmetry which produces the superconducting
phase coherence along the magnetic field.
In the liquid, longitudinal superconductivity is
destroyed by the wandering and entanglement of the vortex lines.
Even though line wandering is energetically costly
and therefore rare, it does occur. As a result, the correlation
length along the c--axis will be quite long and of order $\ell_{z}$.
This is consistent with measurements
in YBCO of the c--axis resistivity which find that there is loss of
vortex velocity correlations
for samples thicker than 100 $\mu$m \cite{Lopez96,Lopez96a,Lopez96b}.
For an infinitely thick sample, the loss of longitudinal
superconductivity coincides with the melting transition \cite{Chen95}. 
This agrees with experiments which indicate that the
loss of superconducting phase coherence along the c--axis
coincides with the first order transition \cite{Lopez96,Lopez96a,Lopez96b}.

\section{\bf Surface Tension}
The vortex line wandering renormalizes the coupling between the planes
in the liquid phase,
so it is difficult to estimate the surface tension parallel to the
$ab$--planes which is primarily due to Josephson vortices. 
However, since we have expressions for the free energy in
both the liquid and solid phases, 
we can estimate the surface tension parallel to the c--axis along the
melting curve.
We imagine a plane interface parallel to the c--axis between the
vortex liquid and the vortex lattice phases. The surface tension 
$\sigma$ is given by  
\begin{equation}
\sigma=\int^{\infty}_{-\infty}(G(x)-G_o)dx
\end{equation}
where $G(x)$ is the Gibbs free energy as a function of position and
the constant $G_o$ is the Gibbs free energy far away from the interface,
for example within the solid phase where $n=0$. 
(At melting the vortex liquid and solid phases coexist because they 
have the same bulk value for the Gibbs free energy.) In the interface region
the defect concentration $n$ changes from zero in the solid phase to
a finite value in the liquid phase. Let us assume that in this 
region the concentration
gradient $dn/dx$ is a constant $n_o/a_o$ where $n_o$ is
the concentration of defects in the bulk 
liquid phase. Here we are assuming that the
width of the interface is of order a vortex lattice constant $a_o$. Then
\begin{equation}
\sigma=\frac{a_o}{n_o}\int^{n_o}_{0}(G(n)-G_o)dn
\end{equation}
This is an integral of the area under the barrier between the solid
and liquid phases in the plot of the Gibbs free energy versus defect
concentration (see inset of Fig. \ref{fig:melting}).
Using the values for $T$ and $B$ along the melting curve, we find
the surface tension given in Fig. \ref{fig:surface}. The dependence
of the surface tension on the melting temperature $T_m$ reflects
that of the barrier height $V_B$ on $T_m$. The order of
magnitude of the surface tension is given by 
$\sigma\sim V_B/sa_o$, and as a result, the values are
quite small. For example, at $T$=60.24 K and $B$=202.28 G, 
$\sigma$=0.015 K/$sa_o$ for BSCCO, and at $T$=80.9184 K and 
$H$=6.4807 T, $\sigma=7.26\times 10^{-3} K/sa_o$ for YBCO, 
where $s$ is the interplane spacing and $a_o=\sqrt{\phi_o/B}$. 
We believe these are correct order of magnitude estimates for the
surface tension since the small values of the barrier height is
consistent with the small amount of hysteresis found experimentally
\cite{Safar93,Zeldov95,Keener97}.

\section{\bf Vortex Liquid}
\subsection{\bf Viscoelastic Behavior}
We now discuss the viscoelastic behavior of the vortex liquid.
As we mentioned earlier, the shear modulus is frequency dependent.
The low frequency response to a shear stress is flow and this 
is characterized by a viscosity $\eta$. At high frequencies the
response is elastic and the crossover between the two occurs over a narrow
frequency range so that the shear modulus as a function of frequency
is rather like a step function. This behavior can be simply modeled using
the Maxwell model \cite{Hansen86} for viscoelasticity in which a massless
spring is damped by a viscous force. The rate of shear strain
$\dot{\varepsilon}$ is given by
\begin{equation}
\dot{\varepsilon}=\frac{\dot{\sigma}}{c_{66}(\omega=\infty)}+\frac{\sigma}
{\eta}
\end{equation}
where $\sigma$ is the shear stress, $\dot{\sigma}$ is the
time derivative of the shear stress, and $\omega$ is the
frequency. Using the Maxwell relation
for the relaxation time $\tau=\eta/c_{66}(\omega=\infty)$ and the
definition of the frequency dependent shear modulus 
$c_{66}(\omega)=\sigma(\omega)/\varepsilon(\omega)$, we find
that the real part of the shear modulus is given by 
\begin{equation}
c_{66}(\omega)=c_{66}(\omega=\infty)\left[\frac{\omega^2\tau^2}
{1+\omega^2\tau^2}\right]
\label{eq:shearmod}
\end{equation}
Notice that $c_{66}(\omega=0)=0$ which confirms that the vortex liquid
cannot sustain a shear stress. At high frequencies $c_{66}(\omega)$
is given by $c_{66}(\omega=\infty)$. To estimate the crossover frequency
we need to estimate the viscosity. 

We are interested in the shear viscosity which arises from the interactions
between vortices. There are other sources of viscosity. For example
a single moving vortex line experiences a viscous drag due to 
the normal electrons in the core which produce resistance
when they move with the vortex. This is described by the Bardeen--Stephen
model \cite{Tinkham96}. There is also viscosity which arises from pinning; 
we will ignore
this contribution since we are considering a clean lattice.
We can obtain a simple estimate for the viscosity following the
approach of Dyre, Olsen and Christensen \cite{Dyre96}. Shear flow occurs
when some vortices push past other vortices. The viscosity
$\eta$ is given by
\begin{equation}
\eta=\eta_o\exp\left[\frac{\Delta F(T)}{k_B T}\right]
\label{eq:visc}
\end{equation}
where the prefactor $\eta_o$ is the viscosity of a single noninteracting
vortex line and is given by the Bardeen--Stephen relation \cite{Tinkham96}
$\eta_o\approx \phi_o H_{c2}/\rho_n c^2$ where $\rho_n$ is the
normal state resistivity and $c$ is the speed of light. In (\ref{eq:visc})
$\Delta F(T)$ is the activation energy which is identified with the
work done per vortex pancake in shoving aside the surrounding vortices. 
The elastic energy associated with distorting the vortices is given by
(\ref{eq:elastic}); we identify $\Delta F(T)$ with ${\cal F}_{el}$ at
the maximum distortion ${\bf u}$ produced by the shoving. 
The actual form of the distortion is difficult to determine analytically.
We will assume that the dominant contribution comes from 
tilt and shear; and that there is no change in the density
so that the contribution from the bulk modulus can be ignored.
In BSCCO in the vortex liquid, the planes are decoupled and 
the tilt modulus can be ignored. In YBCO in the liquid the
correlation along the c--axis can be quite long as we discussed
earlier. In this case the distortion involves various
wavevectors; the wavevector dependence of the tilt modulus
$c_{44}$ is such that at small $q$, $c_{44}$ is roughly 
comparable to the high frequency shear modulus
$c_{66}(\omega=\infty)$ and at large $q$, $c_{44}\ll c_{66}(\omega=\infty)$. 
(Since we are considering the
elastic response, it is appropriate to consider the high frequency shear
modulus.) So as a crude estimate
we will assume the displacement is pure shear and write 
\cite{comment:viscosity}
\begin{equation}
\Delta F(T)=c_{66}(\omega=\infty,n,T)V_c
\label{eq:deltaF}
\end{equation}
where $V_c$ is the volume change due to shoving and rearranging vortices.
Since $\Delta F(T)$ is the energy per vortex pancake, $V_c$ is some
fraction $\delta$ of the volume $\upsilon_o$ per vortex pancake, i.e.,
$V_c=\delta\upsilon_o$. $c_{66}(\omega=\infty,n,T)$ is given by
eq. (\ref{eq:c66exp}). In Fig. \ref{fig:visc} we show the reduced
viscosity $\eta/\eta_o$ along the melting line for both YBCO and BSCCO
with $\delta=1$.
As one can see, interactions enhance the viscosity $\eta$ over the 
noninteracting viscosity $\eta_o$ by a factor of 2 or less. This is 
because $\Delta F(T_m)/k_BT_m<1$ along the melting line. 

Using our estimate of the viscosity and eq. (\ref{eq:shearmod}),
we can calculate the frequency
dependence of the shear modulus $c_{66}(\omega)$ as shown in Fig.
\ref{fig:shearmod} for a defect concentration of 5\%. As expected
the shear modulus has the shape of a step function; it
is zero at low frequencies and rises quite sharply to its infinite
frequency value $c_{66}(\omega=\infty)$.
Notice that for BSCCO the crossover frequency is a few MHz and for YBCO
it is a few GHz. Since 1 K corresponds to 20 GHz, this means that 
we made an excellent approximation in setting 
$c_{66}=c_{66}(\omega=\infty)$ in the free energy density $f$ in
eq. (\ref{eq:freeng}).

\subsection{\bf Dislocations}
Any theory that tries to describe melting has to contain two main
ingredients: a satisfactory description of both the solid and the liquid
phases and a mechanism by which the system goes from one to the other. 
The difficulty has always been to describe two phases with wildly
different properties within the same framework.  In the present work, we
have achieve this by viewing the liquid as a solid with a finite
concentration of vacancies and interstitial. Clearly this is an
approximation. If one wishes to describe a liquid as a solid with
defects, other types of defects have to be taken into account as well.
This is particularly true of dislocations. While we do not believe that
thermally excited
dislocations play the key role in mediating the melting transition
because of their high energy cost, we believe that they will 
proliferate as soon as the vacancies
and interstitials appear. There are two reasons for this.
First interstitial vortex lines are usually attractive
\cite{Olive98} and can aggregate to form dislocations that extend
the entire length of the lattice parallel to the c--axis. The same is true
for vacancies. In particular, Olive and Brandt \cite{Olive98}
have done numerical simulations on line defects which were at least
5 lattice spacings apart. They found that both centered and
edge interstitials \cite{comment:viscosity}
are attractive if $\lambda_{ab}/a_{o}\geq 1$.
For $\lambda_{ab}/a_{o}=0.25$, edge interstitials were attractive
for separations less that 10 lattice spacings and repulsive at larger
distances while centered interstitials were repulsive at distances
larger than 5 lattices spacings. Vacancies were attractive in all cases.

The second reason is that the substantial
softening of the shear modulus brought about by
the vacancies and interstitials reduces the dislocation core energy as well as
the elastic energy of creating dislocations. 
For example, the core energy of a z--directed dislocation
goes as $c_{66}b^2$ and the core energy of a screw dislocation
goes as $\sqrt{c_{66}c_{44}}b^2$ where $b$ is the magnitude of the
Burger's vector \cite{Marchetti99}. In addition the long--range interaction
between dislocation loops is mediated by the strain field and depends
on $c_{66}$ \cite{Marchetti99}. 
Once the dislocations have proliferated, the method and results of
Marchetti and Radzihovsky \cite{Marchetti99} can be used
to provide a more detailed and accurate description of the liquid side of
the transition. For example they show that 
when dislocations at all length scales are present,
the shear modulus vanishes
in the long wavelength limit \cite{Marchetti90,Marchetti99}.
Work in this direction is in progress.

\section{\bf Summary}
To summarize we have presented a model for the melting of a vortex lattice
into a vortex liquid. The melting transition
is induced by a few percent of 
vacancy and interstitial vortex lines that soften
the shear modulus and increase the vibrational
entropy. The increased vibrational entropy leads to melting.
We obtain good agreement with the experimentally
measured curve of transition temperature versus field, latent heat, 
and jumps in magnetization
for BSCCO and YBCO. The Lindemann ratio $c_{L}$ is $\sim 11$\% for BSCCO
and $\sim 25$\% for YBCO. The hysteresis
is small. We find a very small surface tension between the
vortex solid and the vortex liquid along an interface parallel to the
c--axis. The shear modulus is frequency dependent; it is zero at
$\omega=0$ and plateaus at higher frequencies to its infinite frequency
value. 

We thank Andy Granato and Lev Bulaevskii 
for helpful discussions. This work was supported in part by ONR grant
N00014-96-1-0905 and by funds provided by the University of
California for the conduct of discretionary research by Los
Alamos National Laboratory.

\begin{Fig}
\epsfbox{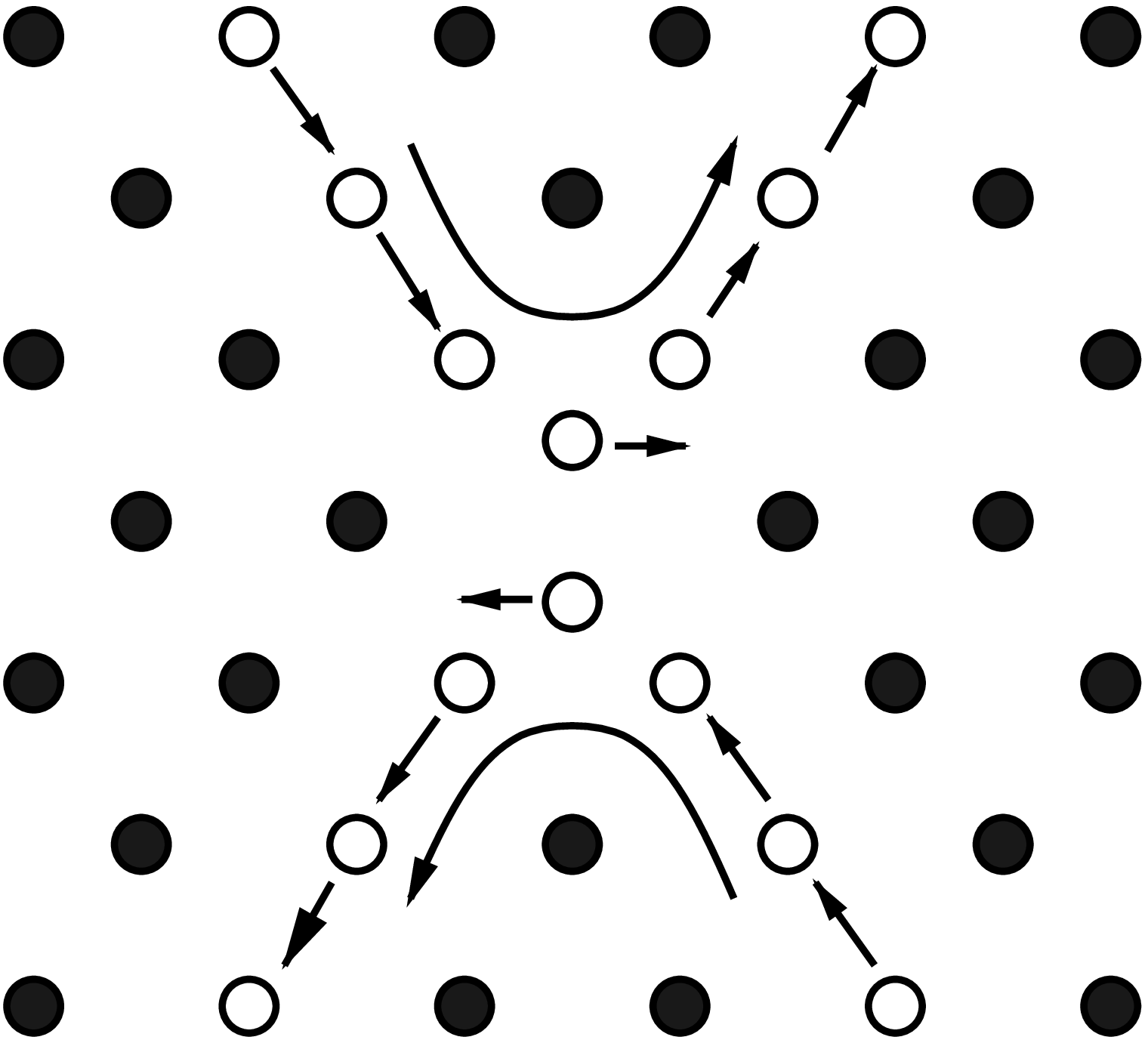}
\caption{Dumbbell interstitialcy configuration in a triangular lattice
introduces a string--like librational
resonance mode that couples to external shear
stress and softens the shear modulus. The atoms or flux lines involved
in this mode are shown with open circles while the rest of the sites
are denoted by solid circles.}
\label{fig:dumbbell}
\end{Fig}
\begin{Fig}
\epsfbox{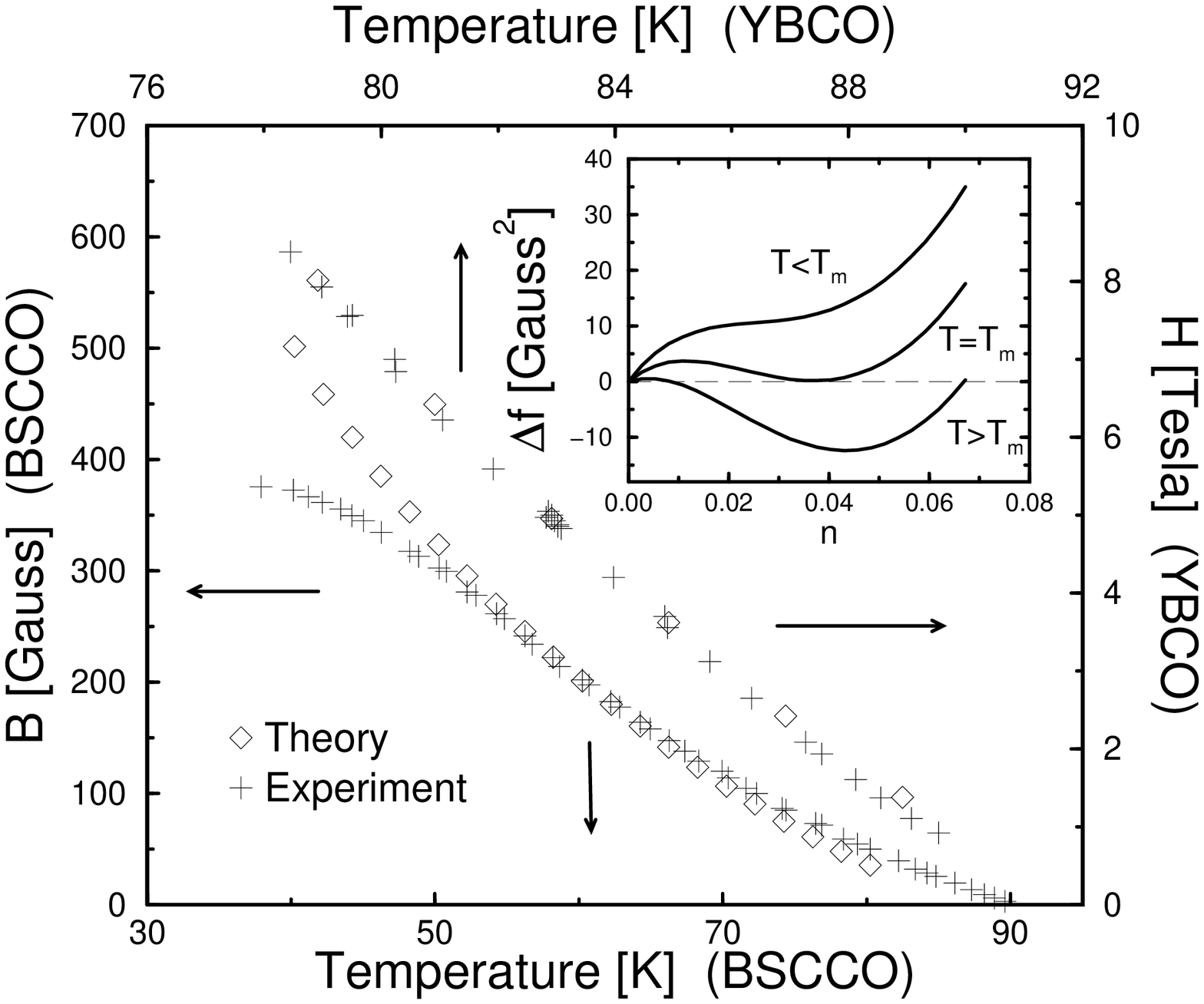}
\caption{First order phase transition curves of 
magnetic field versus temperature.
for YBCO and BSCCO. Parameters used for YBCO are $\alpha_{1}=2.55$,
$\alpha_{2}=0.01485$, $\phi=44.1^{o}$, $\lambda_{ab}(0)=1186 \AA$ 
\protect\cite{Kamal94},
$s=12 \AA$, $\xi_{ab}(0)=15 \AA$, $\gamma=5$, and $T_{c}=92.74$ K.
Parameters used for BSCCO are $\alpha_{1}=1.0$, $\alpha_{2}=0.00705$,
$\phi=60 ^{o}$, $\lambda_{ab}(0)=2000 \AA$, $s=14 \AA$, 
$\xi_{ab}(0)=30 \AA$,
$\gamma=200$, and $T_{c}=90$ K. For BSCCO we use the
low field form of the elastic moduli from (\protect\ref{eq:moduli})
and for YBCO we use the
high field form. For $f_{o}$ we use (\protect\ref{eq:london})
for BSCCO and (\protect\ref{eq:abrikosov}) for YBCO.
(For BSCCO we plot $B$ vs. $T$ because
that is what ref. \protect\cite{Zeldov95} measured.) 
The experimental points
for YBCO come from ref. \protect\cite{Schilling96} and those for BSCCO come
from ref. \protect\cite{Zeldov95}. Inset: Typical $\Delta f$ versus $n$.}
\label{fig:melting} 
\end{Fig}

\begin{Fig}
\epsfbox{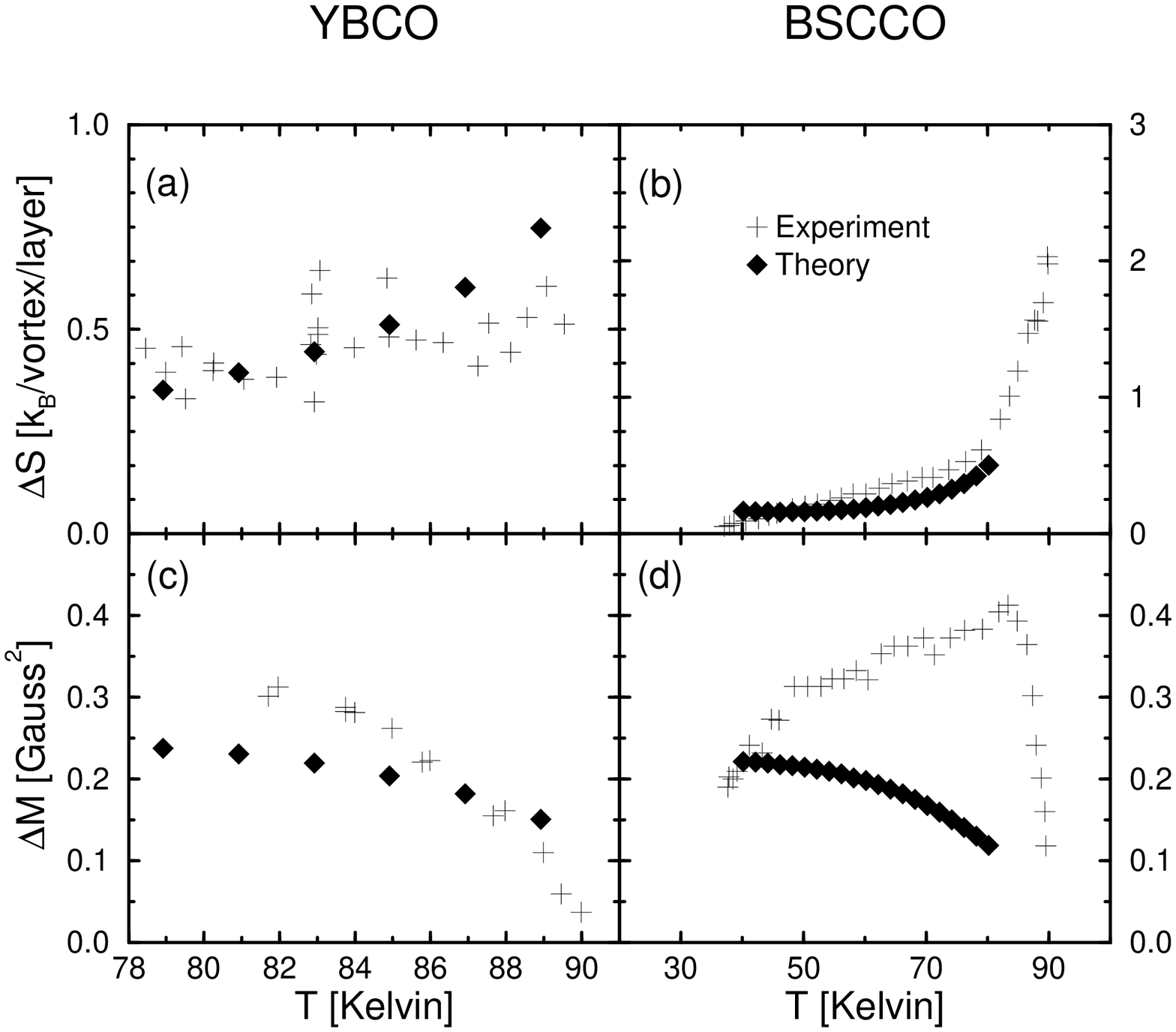}
\caption{(a) and (b): Entropy jump $\Delta s$ per vortex per layer
versus $T_{m}$ at the transition
for YBCO and BSCCO. The experimental points for YBCO are from
\protect\cite{Schilling96} and
those for BSCCO are from \protect\cite{Zeldov95}.
(c) and (d): Magnetization jump $\Delta M$ versus $T_{m}$ at the first
order phase transition for YBCO and BSCCO. 
The experimental points for YBCO are from \protect\cite{Welp96} and 
those for BSCCO are from \protect\cite{Zeldov95}. For the
theoretical points the values of the 
parameters are the same as in Figure 1 for all the curves.}
\label{fig:mag-S}
\end{Fig}

\begin{Fig}
\epsfbox{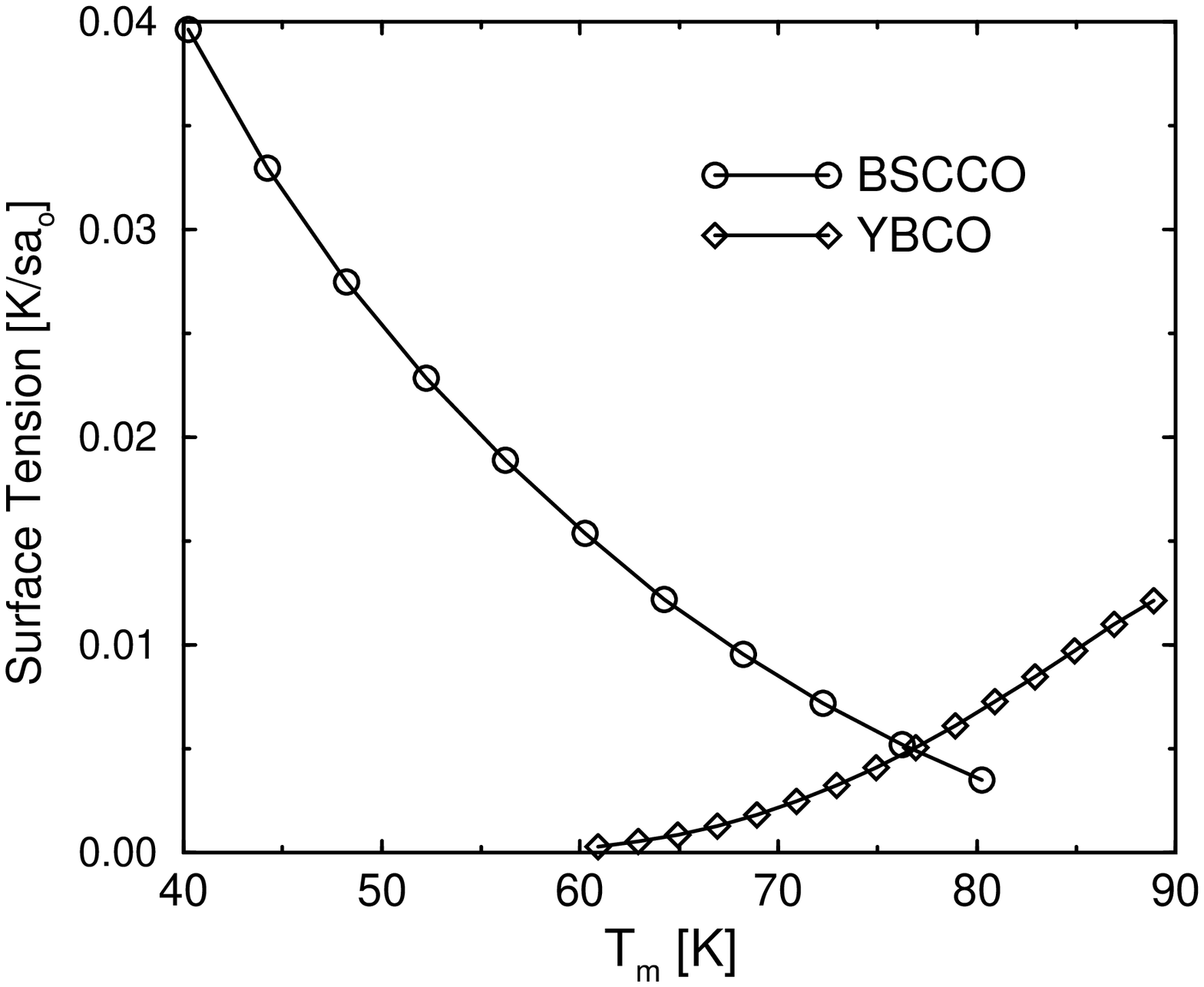}
\caption{Surface tension between the vortex solid and 
liquid phases along an interface parallel to the
c--axis versus melting temperature $T_m$. We use the values
of field and temperature along the melting curve for YBCO and BSCCO.
The surface tension is measured in units of Kelvin/$sa_o$, where
$s$ is the spacing between layers and $a_o=\sqrt{\phi_o/B}$ is the
vortex lattice spacing. The values of the
parameters are the same as in Figure 1.}
\label{fig:surface}
\end{Fig}

\begin{Fig}
\epsfbox{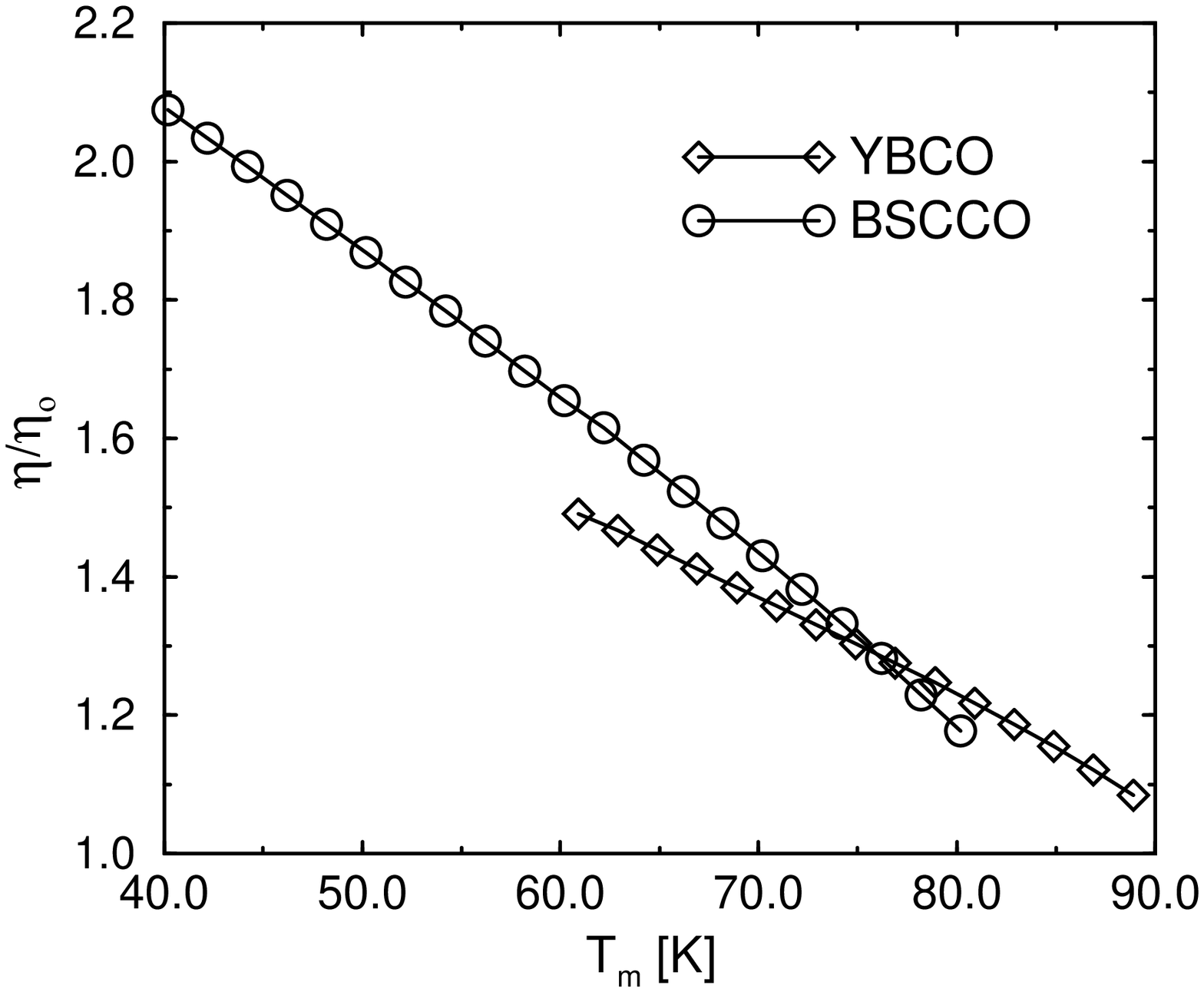}
\caption{Reduced viscosity $\eta/\eta_o$ of the vortex liquid
versus melting temperature $T_m$ along the melting curve
for YBCO and BSCCO. $\delta = 1$. The values of the
parameters are the same as in Figure 1.}
\label{fig:visc}
\end{Fig}

\begin{Fig}
\epsfbox{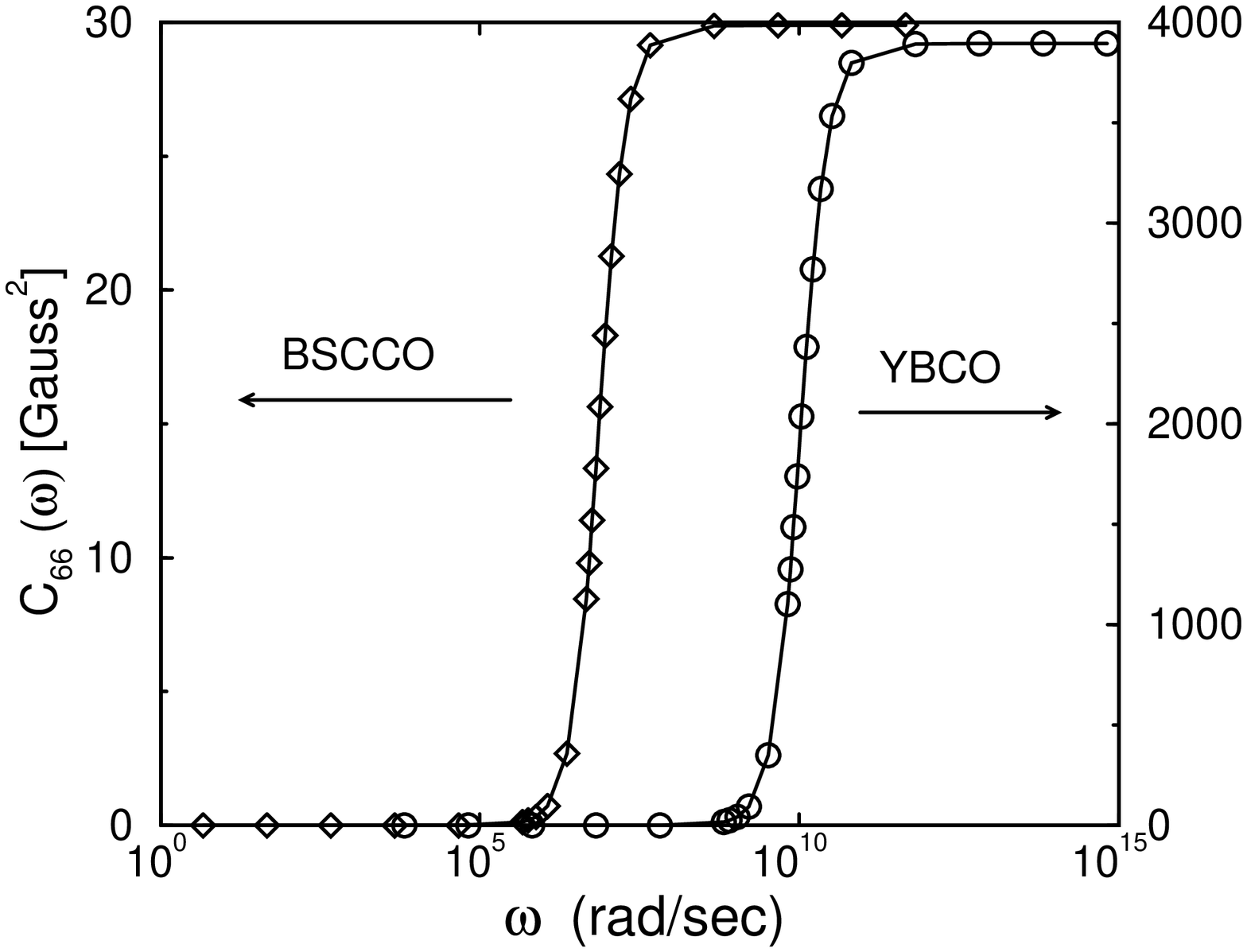}
\caption{Shear modulus $c_{66}$ versus frequency for YBCO and BSCCO.
For YBCO, $H=5$ T, $T=83$ K and the defect concentration $n=4$ \%.
For BSCCO, $B=200$ G, $T=60$ K, and $n=5$ \%. The values chosen are
close to those on the melting curve. The rest of the values of the
parameters are the same as in Figure 1.} 
\label{fig:shearmod}
\end{Fig}

\end{multicols}

\begin{thebibliography}{10}
\bibitem{comment:marchetti}
Although calculations indicate that fluctuations in the density of vacancies
  and interstitial defects do not renormalize $c_{66}$
  \protect\cite{Marchetti99}, $c_{66}$ does decrease as the average number of
  defects increases.
\bibitem{comment:phi}
Our $\phi$ equals half the angle cited in the experiments.
\bibitem{comment:viscosity}
Another way to estimate the free energy $\Delta F(T)$ in
  (\protect\ref{eq:visc}) is to use the energy difference between a centered
  interstitial and an edge interstitial \protect\cite{Frey94,Olive98}. A
  centered interstitial vortex line is at the center of a triangle whose
  vertices are the nearest neighbor lattice sites of a triangular lattice. An
  edge interstitial lies at the midpoint of one of the sides of the triangle.
  Numerical calculations \protect\cite{Frey94,Olive98} find this energy to be
  roughly $10^{-3}\cdot 2\epsilon_{o}\sim 1$ K $\ll T_{m}$. So on the melting
  curve, $\eta/\eta_{o}\simeq 1$. This estimate of $\Delta F(T)$ is an order of
  magnitude smaller than the estimate we used in eq. (\protect\ref{eq:deltaF}).
  As a result the the crossover frequency, where $c_{66}(\omega)$ goes from its
  low frequency to its high frequency value, increases by less than a factor of
  two compared to the crossover frequencies shown in Fig.
  \protect\ref{fig:shearmod}.
\end{thebibliography}
\end{document}